\documentclass[twocolumn,apl,reprint,showpacs,preprintnumbers,amsmath,amssymb,superscriptaddress]{revtex4-2}
\usepackage{graphicx}
\usepackage{dcolumn}
\usepackage{bm}
\usepackage{amsmath}
\usepackage{amssymb}
\usepackage{braket}
\usepackage{todonotes}
\usepackage{color}
\usepackage{bbm}
\usepackage{mathtools}
\usepackage{epsfig}
\usepackage{epstopdf}
\usepackage{color}
\usepackage[english]{babel}
\usepackage{bm}
\usepackage{subfigure}
\usepackage{hyperref}
\usepackage{xcolor}
\usepackage[autostyle]{csquotes}

\usepackage[normalem]{ulem}
\hypersetup{
    colorlinks,
    citecolor=blue,
    filecolor=black,
    linkcolor=blue,
    urlcolor=blue
}
 \usepackage{relsize}

\newcommand*{\rom}[1]{\expandafter\@slowromancap\romannumeral #1@}
\begin{document}


\title{Brownian dynamics of Dirac fermions in twisted bilayer graphene}

\author{Abdullah Yar*}
\affiliation{Department of Physics, Kohat University of Science and Technology,\\
Kohat 26000, Khyber Pakhtunkhwa, Pakistan}

%
\email{abdullahyardawar@gmail.com}

\begin{abstract}
Brownian dynamics of Dirac fermions in twisted bilayer graphene is
investigated within the framework of semiclassical relativistic
Langevin equations.
We find that under the influence
of orthogonal, commensurate ac drives in the
periodic ratchet potential of a substrate, the charge carriers
in the system exhibit pronounced
random dynamics, tuned by the twist angle, making twisted bilayer graphene
distinguishable from monolayer graphene.
It is shown that as threshold twist angle matches the optimal angle, deterministic
running states appear in the
limit of weak thermal noise where the diffusion rate is enhanced
significantly compared to bare thermal diffusion.
Analysis of the real space trajectories and diffusion
coefficient illustrates the significant role of
thermal noise in the random motion of Dirac fermions.
In addition, we find that the
Brownian particle shows remarkable ratchet effect as a net current.
\end{abstract}


\maketitle

\section{Introduction}\label{Introduction}
The study of exotic properties of twisted bilayer graphene (TBG)
has attracted significant attention in condensed matter
physics~\cite{Cao-Nat.556:80,Cao-Nat.556:43,Yankowitz-SC.363:1059,Koshino-PRX.8:031087,Tarnopolsky-PRL.122:106405,Liu-PRB.99:155415,Song-PRL.123:036401,Kang-PRX.8:031088}. It is composed of two monolayer graphene sheets, obtained by
stacking one layer of graphene on top of another and rotating
the two layers by a small twist angle, leading to a moir\'{e}
pattern to arise between the lattices of the two graphene layers.
It can be realized as a model platform for investigating the properties
of semiconducting flat band moir\'{e} van der Waals heterostructures.
It has been found that the Dirac crossings in TBG are protected by the product of two-fold
rotational and time-reversal symmetries ($C_2T$)~\cite{Po-PRX.8:031089}.
A long period moir\'{e} pattern is formed if stacking of the two graphene
monolayers is slightly disoriented, leading to a significant reduction of
Fermi velocity at the Dirac cone. In TBG, the concentration of charge
density is maximal around a moir\'{e} pattern, leading to the formation of a triangular lattice structure~\cite{Li-Nat.6:109,Laissardiere-NL.10:804,Luican-PRL.106:126802,Wong-PRB.92:155409,Kim-PNAS.114:3364}.
The quantum states of Dirac fermions in the vicinity of each valley
of graphene sheet hybridize with the electronic states
of the other sheet, resulting into interesting electronic band structure.
It has been shown that two nearly flat bands are
formed in the middle of the full energy band, separated from other energy bands
for the twisting angle being close to magic angles~\cite{Bistritzer-PNAS.108:12233}.
The interaction effects are very enhanced when the chemical
potential remains inside these nearly flat bands. Remarkable
unconventional superconducting phases~\cite{Cao-Nat.556:43}
and Mott insulator behavior~\cite{Cao-Nat.556:80}
have been observed in the transport studied of TBG.
Interesting dissipation processes~\cite{Fang-PRB.93:235153,Santos-PRL.99:256802,Jose-PRL.108:216802,Mele-PRB.81:161405}, temperature-dependent electrical
transport~\cite{He-NP.17:26}, tunable large Berry
curvature~\cite{Pantaleon-PRB.103:205403}, nonlinear
Hall effect~\cite{Zhang-PRB.106:L041111}, carrier
confinement~\cite{Tilak-NC.12:4180}, and topological
Hofstadter butterfly~\cite{Lu-PNAS.118:2100006118}
have been investigated in TBG.
The flat bands in twisted bilayer graphene have been
observed experimentally using local tunnelling spectroscopy technique~\cite{Kerelsky-Nat.572:95,Xie-Nat.572:101,Jiang-Nat.573:91,Li-Nat.6:109,Choi-Nat.15:1174},
electronic compressibility measurements~\cite{Tomarken-PRL.123:046601},
and direct momentum-resolved measurements~\cite{Utama-Nat.17:184}.
Another possible method for detecting flat bands in
TBG is the combination of different imaging techniques
with angle-resolved photoemission spectroscopy (ARPES) that can be used to observe
the flat band structures in twisted bilayer graphene devices~\cite{Lisi-Nat.17:189}.\\
On the other hand, the study of Brownian motion has been the focus
of active research~\cite{Haenggi-RMP.81:387}. It is the description
of any physical phenomenon in which some quantities are constantly
undergoing small random fluctuations.
It also provides insights into the numerical solution of stochastic
differential equations.
The Brownian particle experiences
irregular forces as a result of random collisions with atoms or
molecules of the medium, which alter both the direction and magnitude
of the velocity of charge carrier. The study of Brownian motion has played an
important role in the development of both the foundations
of thermodynamics and the dynamical interpretation of statistical physics.
The theory of Brownian motion based on molecular-kinetic theory of
heat provides the link between an elementary underlying
microscopic dynamics and macroscopic observable phenomena.
It has been shown that a particle undergoes a zero-temperature
localization-delocalization transition in a periodic potential
with the decrease of dissipation strength~\cite{Friedman-PRB.100:060301}.
Moreover, optimal estimation of diffusion coefficient of a
diffusing particle from a single-time-lapse
recorded trajectory of the particle has been studied~\cite{Vestergaard-PRE.89:022726}.
Likewise, periodically driven Brownian motion of a massive particle in one
dimension subjected to dry friction has been investigated~\cite{Pototsky-PRE.87:32132}.
Recently, non-random dynamics of DNA molecules in the regime
where length scales are greater than the size of molecule,
undetectable by the analysis of mean squared displacement (MSD), a
measure of the deviation of the position of a particle with
respect to a reference position, has been quantified by characterizing
the molecular motion relative to a latticed frame of reference~\cite{Serag-NC.8:15675}.
Moreover, other interesting aspects of Brownian
motion have been studied. For instance, active random motion
with intermittent direction reversals~\cite{Santra-PRE.104:L012601},
nanoparticle dynamics in nanofluid~\cite{Hajaj-P.9:1965},
highly resolved random motion in space and time~\cite{Mo-ARFM.51:403},
Brownian behavior in coupled chaotic oscillators~\cite{Pasquin-M.9:2503}, paradigmatic model of Brownian motion in a harmonic oscillator coupled to a bath of oscillators~\cite{Boyanovsky-PRA.96:062108}, Brownian motion in a Landau level system~\cite{Cobanera-PRB.93:245422} have been reported. \\
It has been shown that a driven Brownian dynamics is described
by two states: (i) locked state,
where the Brownian particle resides inside one potential well, which
is obtained in the limit of small driving force strength.
(ii) Running state in which the particle runs over the potential
barriers, which appears in the large driving force strength regime.
In these regions, both the diffusion and regular behaviours of the
Brownian particle are accessible.
Brownian dynamics can be characterized by the standards: (i) availability of noise, where
the interplay among nonlinearity, noise-mediated dynamics, and non-equilibrium
driving leads to an unpredictable electronic transport. (ii) Temporal periodically
driven supplemented symmetry breaking, usually involved
in the periodically operating devices.\\
In view of the promising characteristic features exhibited
by the Dirac fermions, we report the Brownian dynamics in
twisted bilayer graphene (TBG). It has been found that Dirac fermions in TBG show
transport properties different from monolayer graphene, for instance, the
TBG exhibit finite nonlinear Hall effect~\cite{Zhang-PRB.106:L041111}
which vanishes in monolayer graphene.\\
In this work, we show that the Dirac fermions in TBG exhibit pronounced Brownian
motion, tuned by the twist angle, presenting transport
scenario different from monolayer graphene.
In addition, we investigate ratchet effect as a net current.
The investigation of Brownian dynamics in TBG may be an
interesting aspect of dynamics with the possible experimental observation.\\
The paper is organized as follows: Methodology for the Brownian dynamics of Dirac
fermions in twisted bilayer graphene is presented in Sec.~\ref{Sec:Methodology}.
The band structure of TBG is determined, followed by the description of
relativistic semiclassical Langevin equations.
In Sec.~\ref{Sec:RD}, results and discussion are presented with the evaluation and
analysis of diffusion coefficient as a function of thermal noise,
ratchet potential strength, driving force field amplitude, and twist angle.
Ratcheting as a net current is also demonstrated in this section.
Finally, the summary of results is presented in Sec.~\ref{Sec:Conc}.
\section{Methodology} \label{Sec:Methodology}
In this section, we present the detailed methodology of the
diffusion mechanism in TBG driven by orthogonal,
commensurate ac drives in a periodic ratchet potential of a substrate.
\subsection{Model Hamiltonian} \label{Sec:MH}
We consider a hexagonal-lattice model that reflects the symmetry
properties and number of subbands of the flat band in twisted bilayer
graphene (TBG).
Taking into account the nearest neighbor hopping, the tight-binding
Hamiltonian within the continuum model approach can
be expressed as~\cite{Lewandowski-PNAS.116:20869}
\begin{align}\label{eq:Hamiltonian}
\mathcal{H}_0\left({\bf{k}}\right) &=\left( \begin{array}{cc} 0 & h_{\bf{k}} \\
h^\ast_{\bf{k}} & 0 \end{array} \right),
\end{align}%
with
\begin{align}\label{eq:hopHamiltonian}
h_{\bf{k}}=\frac{W}{3}\sum_{\pmb{\delta}_l}e^{i{\bf{k}}\cdot\pmb{\delta}_l},
\end{align}%
where $W/3$ represents the hopping matrix element to nearest neighbours at
positions $\pmb{\delta}_l=\left(\cos\left(2\pi l/3\right),\sin\left(2\pi l/3\right)\right)L_M/\sqrt{3}$, where $l=0,1,2$ and $L_M$ is the
moir\'{e} superlattice periodicity, realized in twisted bilayer graphene.
It is illustrated that $W$ represents the bandwidth measured from Dirac point,
and the nearest neighbour distance $L_M/\sqrt{3}$ is chosen
such that the lattice period of the hexagonal model mimics the moir\'{e} superlattice period.
Following the analysis of Lewandowski and Levitov~\cite{Lewandowski-PNAS.116:20869},
we choose the energy and length scales such as the width of a single band $W$ and the hexagonal
lattice period $L_M$ to be identical to the parameters in
TBG: $W = 3.75 \textrm{meV}$ and $L_M = a/2 \sin\left(\theta/2\right)$ is the
moir\'{e} superlattice period, yielding $L_M = 13.4 \textrm{nm}$ for the magic
angle $\theta = 1.05^{\circ}$ and lattice constant of graphene, $a = 0.246 \textrm{nm}$.
The relativistic energy dispersion of the Dirac fermions is
\begin{align}\label{eq:Spectrum}
 E_\lambda\left(\bf{k}\right)=\lambda |h_{\bf{k}}|,
\end{align}%
where $\lambda=\pm$ is the band index and $|h_{\bf{k}}|=\frac{W}{3}
\sqrt{1+4\cos\left(\frac{L_M}{2}k_y\right)
\cos\left(\frac{\sqrt{3}L_M}{2}k_x\right)
+4\cos^2\left(\frac{L_M}{2}k_y\right)}$.
In the following analysis, we use Eq.~\eqref{eq:Spectrum} with $\lambda=+$.
\subsection{Relativistic Langevin equations}\label{Sec:LE}
We consider a sample of twisted bilayer graphene (TBG) on a substrate
with a periodic potential $V(x)$. TBG is driven by two orthogonal
harmonic drives $E_x(t)$ and $E_y(t)$ with commensurate frequencies,
inducing rectification that leads to a direct current.
The ratchet substrate in one dimension is modeled by the double-sine
potential as~\cite{Haenggi-RMP.81:387,Pototsky-EPJB.85:356,Derivaux-JSM.043203,Reimann-PRL.79:10}
\begin{align}\label{eq:pot}
V\left(x\right)=V_0\left[\sin\left(\frac{2\pi x}{L}\right)
+\mu \sin\left(\frac{4\pi x}{L}\right)\right],
\end{align}
where $V_0$ is the strength, $L$ is the period of the potential,
and $\mu=1/4$ is a constant.
The Brownian motion of Dirac fermions in TBG can be described by a set of
coupled relativistic Langevin equations for the components
of 2D momentum, ${\bf{k}} = \left(k_x, k_y\right)$, subjected
to the relativistic dispersion relation given in Eq.~\eqref{eq:Spectrum}.
These equations are determined by the condition that the chosen
particle-reservoir coupling must lead to the same equilibrium
momentum distribution as predicted by the fully microscopic theory.
We follow the procedure of Dunkel {\it et al.}~\cite{Dunkel-PR.471:1} and
Pototsky {\it et al.}~\cite{Pototsky-EPJB.85:356}
for deriving the coupled relativistic Langevin equations.
For describing the random dynamics, charge carriers in TBG require high
energies for their distribution to be appropriately approximated by the
relativistic J\"{u}ttner distribution,
\begin{align}\label{eq:LE1}
f\left(\bf{k}\right)\sim \exp\left[-E\left(\bf{k}\right)/k_BT\right].
\end{align}
For a technical discussion on validity of the relativistic
J\"{u}ttner distribution, the relevant approximations and detailed
derivations of the coupled relativistic Langevin
equations, the reader is referred to Refs.~\cite{Dunkel-PR.471:1,Pototsky-EPJB.85:356}.
Here we emphasize that the J\"{u}ttner
distribution can be regarded as a semiclassical approximation
of the Fermi-Dirac distribution for relativistic particles with rest energy
on the order of, or smaller than $k_BT$.
Such a condition is valid for charge carriers in TBG in
the vicinity of Dirac points. In cartesian coordinates, a
convenient~\cite{Pototsky-EPJB.85:356} but not unique~\cite{Dunkel-PR.471:1}
set of semiclassical coupled relativistic Langevin equations consistent with
the above 2D relativistic J\"{u}ttner distribution is
\begin{align}\label{eq:LE1}
&\dot{x}=\frac{1}{\hbar}\frac{\partial |h_{\bf{k}}|}{\partial k_x},
\end{align}
\begin{align}\label{eq:LE2}
&\dot{y}=\frac{1}{\hbar}\frac{\partial |h_{\bf{k}}|}{\partial k_y},
\end{align}
\begin{align}\label{eq:LE3}
&\dot{k}_x =-\frac{\eta}{\hbar^2}\frac{\partial |h_{\bf{k}}|}{\partial k_x}
-\frac{1}{\hbar}\frac{dV(x)}{dx}
+\frac{e}{\hbar}E_x\left(t\right)+\frac{1}{\hbar}\sqrt{2\eta k_BT}\zeta_x\left(t\right),
\end{align}
\begin{align}\label{eq:LE4}
&\dot{k}_y =-\frac{\eta}{\hbar^2}\frac{\partial |h_{\bf{k}}|}{\partial k_y}
+\frac{e}{\hbar}E_y\left(t\right)
+\frac{1}{\hbar}\sqrt{2\eta k_BT}\zeta_y\left(t\right),
\end{align}
where $\eta$ is the damping constant, $E_x\left(t\right)$ and
$E_y\left(t\right)$ are the driving ac electric fields in the $x$- and
$y$-directions, respectively, whereas
\begin{align}\label{eq:Der1}
\frac{\partial |h_{\bf{k}}|}{\partial k_x}&=-\frac{WL_M}{\sqrt{3}}\times\notag\\&\frac{\cos\left(\frac{L_M}{2}k_y\right)
\sin\left(\frac{\sqrt{3}L_M}{2}k_x\right)}{\sqrt{1+4\cos\left(\frac{L_M}{2}k_y\right)
\cos\left(\frac{\sqrt{3}L_M}{2}k_x\right)+4\cos^2\left(\frac{L_M}{2}k_y\right)}},
\end{align}
and
\begin{align}\label{eq:Der2}
\frac{\partial |h_{\bf{k}}|}{\partial k_y}&=-\frac{WL_M}{3\hbar}\times\notag\\&\frac{\left[\cos\left(\frac{\sqrt{3}L_M}{2}k_x\right)
\sin\left(\frac{L_M}{2}k_y\right)+\sin\left(L_Mk_y\right)\right]}{\sqrt{1+4\cos\left(\frac{L_M}{2}k_y\right)
\cos\left(\frac{\sqrt{3}L_M}{2}k_x\right)+4\cos^2\left(\frac{L_M}{2}k_y\right)}}.
\end{align}
The random forces
$\zeta_x\left(t\right)$ and $\zeta_y\left(t\right)$ in Eqs.~\eqref{eq:LE3} and~\eqref{eq:LE4} are two white
Gaussian noises with vanishing mean, $\braket{\xi_m\left(t\right)}=0$, satisfying
the fluctuation-dissipation relation,
$\braket{\zeta_m\left(t\right)\zeta_n\left(t_0\right)}=
\delta_{mn}\delta\left(t-t_0\right)$, with $m,n=x,y$, which ensures
proper thermalization at temperature $T$.
The damping coefficient $\gamma$ contains contributions
from all possible collision mechanisms including phonon scattering, a charge
carrier may undergo when moving on the surface of a TI attached to SMO.
In order to take into account microscopic collision mechanisms
requires a detailed quantum kinetic theory,
for instance, see~\cite{Mulle-PRL.103:025301}, which is beyond the scope of
our semi-classical approach considered in this paper.
Note that white Gaussian noise works appropriately only for sufficiently high temperatures.
Analysis of Eqs.~\eqref{eq:LE1}-~\eqref{eq:LE4} reveals that
the orthogonal ac-drive components $E_x(t)$ and $E_y(t)$ are
nonlinearly coupled through the relativistic energy dispersion, leading to
significant rectification induced by the periodic ratchet potential $V(x)$.
The above mentioned equations describe the
dynamics of a particle under the influence of driving forces and potential
of a substrate which include effective parameters containing $\hbar$.
Moreover, in these dynamical equations, $x$ and $y$ are real-space
variables, whereas $k_x$ and $k_y$ are the $x$- and $y$- components,
respectively, of the crystal momentum which can also be regarded
as expectation values of their respective operators.
It has been shown that two harmonic drives
with commensurate frequencies can induce rectification, leading to direct
current in the non-relativistic regime when both are applied along the same direction~\cite{Marchesoni-PLA.119:221,Savelev-EPL.67:179,Savelev-EPLB.40:403}.
Remarkably, in the relativistic Langevin equations in Eqs.~\eqref{eq:LE1}-~\eqref{eq:LE4},
coupling between the $x$ and $y$ degrees of freedom generates an unusual
harmonic mixing, where rectification of current occurs even if
the two harmonic drives are orthogonal to each other. Moreover,
the damping coefficient $\eta$ takes into account contributions
from all possible collision mechanisms a charge carrier may undergo
when moving through a twisted bilayer graphene sheet.
Furthermore, the noisy environment in TBG can be provided
by external noise sources such as
current or voltage fluctuations. As a consequence, the interaction of a single
charge carrier with such an environment generates damping effects, which are
included in relativistic Langevin equations. The appropriate form of the damping
and the fluctuating terms can be determined by the fluctuation-dissipation theorem.
\section{Results and Discussion}\label{Sec:RD}
In this section, we present the results based on our
semi-classical approach for describing Brownian dynamics of
Dirac fermions in twisted bilayer graphene.
It is illustrated that all simulations involved in the solution of
Eqs.~\eqref{eq:LE1}-~\eqref{eq:LE4} have been performed employing the stochastic Runge-Kutta
algorithm~\cite{Honeycutt-PRA.45:600}.
\subsection{Relativistic ratchet effect}\label{Sec:RR}
In this section, we consider charge carrier dynamics in
the presence of rectification
of non-equilibrium perturbations on a substrate
with periodic ratchet potential $V(x)$.
\begin{figure}[!ht]
\centerline{\psfig{figure=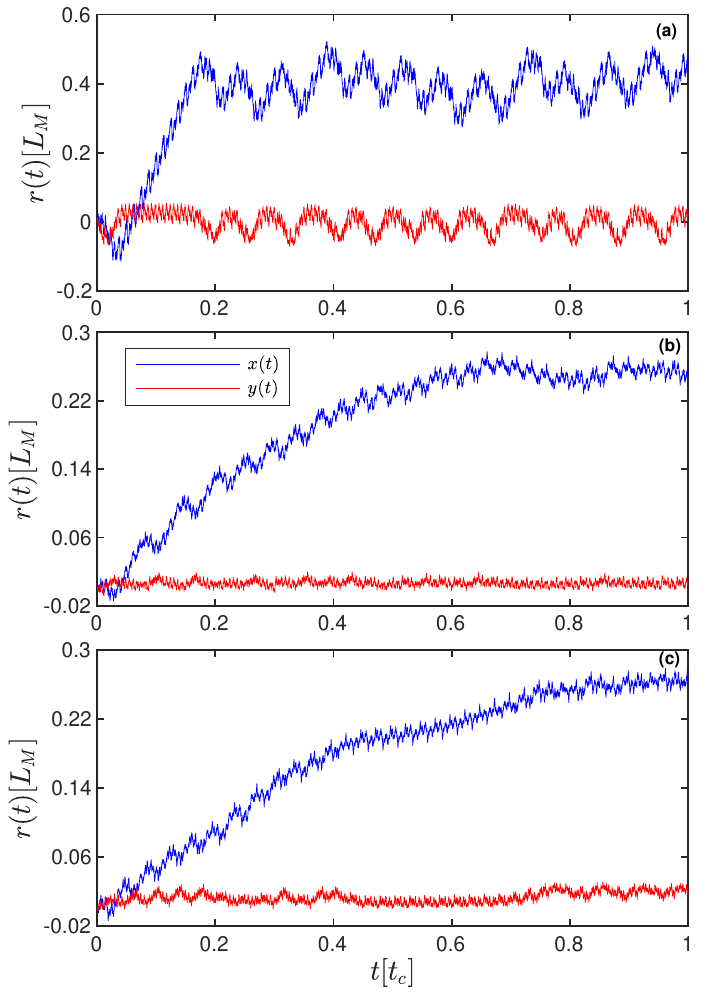,width=\columnwidth}}
\caption{Real space trajectories of Brownian dynamics in
twisted bilayer graphene for the twist angles: (a) $\theta = 1.03^{\circ}$, (b) $\theta = 1.05^{\circ}$, and (c) $\theta = 1.06^{\circ}$ in units of $L_M$ for $\theta = 1.05^{\circ}$. The blue curve shows the
$x$-component and the red curve is used for the $y$-component
of the dynamics. The parameters used are: $t_c=\frac{\hbar}{W}$, $\omega_x =  \omega_0$ with $\omega_0=W/\hbar$,
$\omega_y = 4\omega_x$, $eE_x = F_0$ with $F_0=W/a$, $eE_y =8eE_x$, $\eta =  0.03F_0/v_\textrm{F}$ with $v_\textrm{F}=Wa/\hbar$, $L=700a$, $k_BT=0.001W$, $V_0=120W$.}
\label{Figure1}
\end{figure}
\begin{figure}[!ht]
\centerline{\psfig{figure=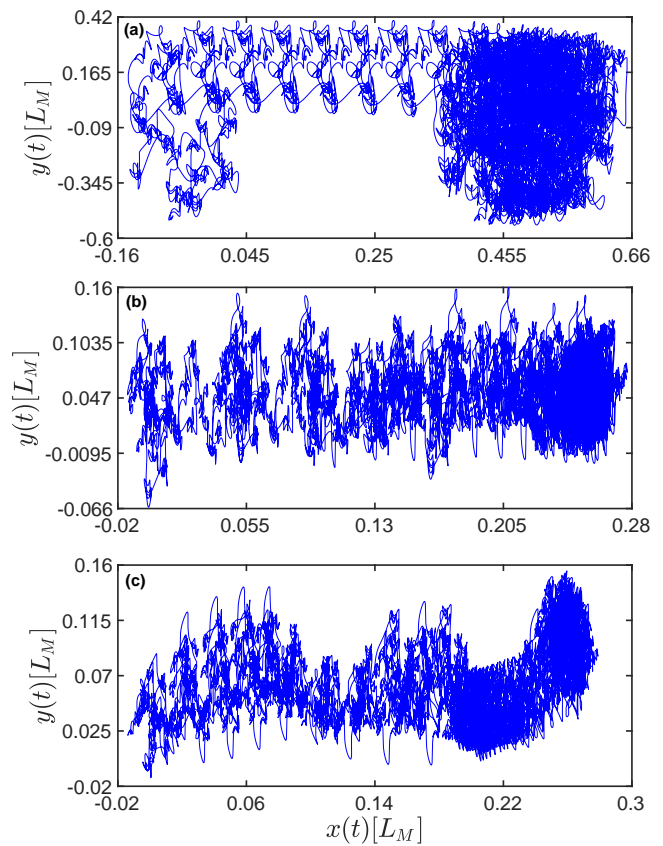,width=\columnwidth}}
\caption{Real space trajectories of Brownian dynamics in
twisted bilayer graphene for the twist angle:
(a) $\theta = 1.03^{\circ}$, (b) $\theta = 1.05^{\circ}$, and
(c) $\theta = 1.06^{\circ}$ in units of $L_M$ for $\theta = 1.05^{\circ}$ in the $xy$-plane using the same
parameters as used for Fig.~\ref{Figure1}.}
\label{Figure2}
\end{figure}
Here, we consider the driving electric field as ${\bf{E}}\left(t\right)=\left(E_x\cos\left(\omega_xt\right),E_y\cos\left(\omega_yt\right)\right)$,
where $E_x$ and $E_y$ are the amplitudes, $\omega_x$ and $\omega_y$ are
the oscillation frequencies of the $x$- and $y$-components of the field.
It has been shown that in the non-relativistic regime, ratcheting takes
place only in the $x$-direction, induced by the
periodic driving field, $E_x(t)$ oriented along $x$-axis. The
rectification weakens if $\textbf{E}(t)$ is rotated at an angle
with the $x$-axis, until it drops to zero for ac drives parallel
to the substrate valleys. In this process, the component $E_y(t)$ of
the ac-field keeps the system out of equilibrium, however, it
cannot be rectified, as the substrate potential is uniform
in the $y$ direction. Hence, the $x$ and $y$ dynamics
remain decoupled. In the relativistic Langevin
equations, see Eqs.~\eqref{eq:LE1}-~\eqref{eq:LE4},
instead, the orthogonal ac-drive components,
$E_x(t)$ and $E_y(t)$, are nonlinearly coupled through
the relativistic dispersion relation, see Eq.~\eqref{eq:Spectrum}, so that both can be rectified by
the periodic ratchet potential $V(x)$.
In order to understand the diffusion mechanism in twisted
bilayer graphene, we analyze the real space trajectories of the
driven Brownian dynamics.
In Fig.~\ref{Figure1}, the real-space trajectories of the mentioned dynamics
have been shown for different values of the twist angle.
Comparison of panels (a), (b), and (c) reveals that the
Brownian dynamics changes significantly with the change of
twist angle.
In particular, the trajectories of Brownian particle
for twist angle, $\theta = 1.03^{\circ}$ are significantly different
from those plotted for $\theta = 1.05^{\circ}$ and $\theta = 1.06^{\circ}$,
showing that the dynamics in TBG strongly depends on the
twist angle and the randomness is more prominent for $0<\theta<1.05^{\circ}$.
Moreover, comparison of the blue and red curves
shows that the $x$- and $y$-components of Brownian dynamics
exhibit different dynamical behavior.
In Fig.~\ref{Figure2}, the real-space trajectories of Brownian particle
have been shown for different values of the twist angle in the $xy$-plane.
This figure also shows that the trajectories of Brownian particle exhibit
irregular behavior.
The characteristic feature of Brownian
dynamics in TBG stems from the fact that the trajectories fall in
the experimentally accessible regime that can be measured
using the recently developed experimental
techniques~\cite{Kheifets-Sc.343:1493,Petrov-SR.12:8618}.
Such trajectories have also been investigated
in various systems~\cite{Ma-JCP.145:114102,Evstigneev-JPA.52:055001}.
However, the amplitudes of trajectories in TGB are
an order of magnitude larger than these systems. Analysis
of Figs.~\ref{Figure1} and~\ref{Figure2} reflects
random behavior of the particle dynamics, hallmark of
Brownian dynamics. For further analysis and understanding the dynamics,
we analyze the diffusion coefficient and net current in the following sections.
\subsubsection{Diffusion coefficient}\label{Sec:DC}
The Brownian particle described by the dynamical equations
in Eqs.~\eqref{eq:LE1}-~\eqref{eq:LE4} drifts with average
speed $\braket{{\dot{\bf{r}}}}$ in the direction of the applied resultant
force, the random switches between locked and running states
lead to a spatial dispersion of the particle around its average position.
\begin{figure}[!ht]
\centerline{\psfig{figure=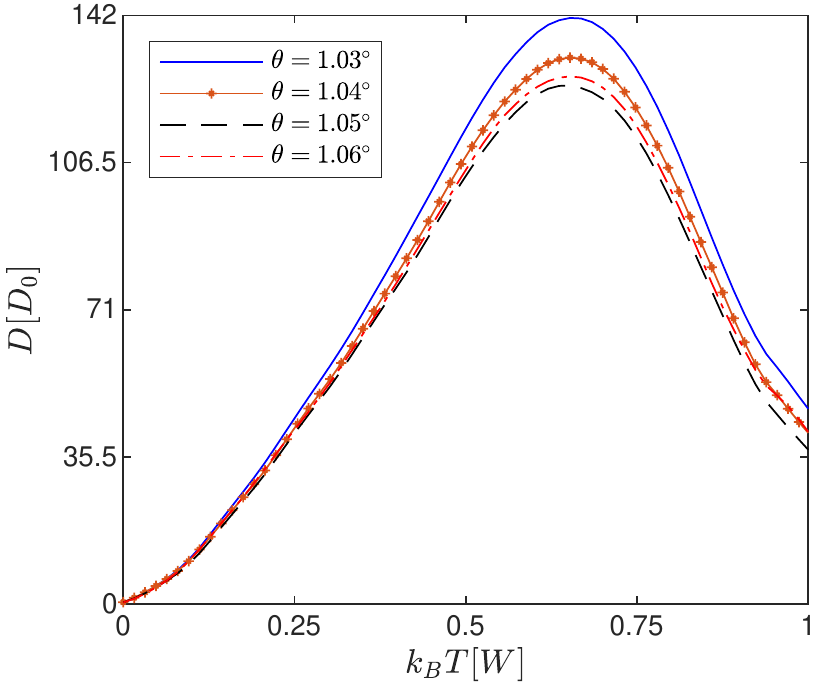,width=\columnwidth}}
\caption{Diffusion coefficient $D$ in units of
$D_0=\frac{k_BT}{\eta}$ versus thermal noise strength $k_BT$. The blue
solid curve shows the diffusion coefficient for $\theta = 1.03^{\circ}$, the red asteric solid curve is used for $\theta = 1.04^{\circ}$, the black dashed curve is used for $\theta = 1.05^{\circ}$, and
the green dash-dotted curve characterizes the diffusion coefficient for $\theta = 1.06^{\circ}$.
The parameters used in numerical simulations are: $\omega_x =  \omega_0$,
$\omega_y = 4\omega_x$, $eE_x = F_0$, $eE_y =7eE_x$, $\eta =  0.0001F_0/v_\textrm{F}$, $L=700a$.}
\label{Figure3}
\end{figure}
In the long-time limit, the corresponding effective diffusion coefficient can be expressed as~\cite{Haenggi-RMP.81:387}
\begin{align}\label{Eq:DC}
D:= \lim_{t\to\infty} \frac{\braket{{\bf{r}}^2\left(t\right)}-\braket{{\bf{r}}\left(t\right)}^2}{2t},
\end{align}
where the two brackets denote average evaluated over the initial
conditions of position and over all realizations of thermal noise and
${\bf{r}}=\left(x,y\right)$ are the $x$- and $y$-components of the trajectory.
Fig.~\ref{Figure3} shows the temperature dependence
of diffusion coefficient in TBG for different values of the twist angle.
This figure shows that
the diffusion coefficient exhibits resonancelike behavior,
where maximal is observed around $k_BT \sim 0.65W$,
revealing the existence of an optimal $k_BT$
for the enhancement of diffusion rate. This mechanism is reminiscent of stochastic resonance~\cite{Haenggi-RMP.62:251,Benzi-JPA.14:453,Gammaitoni-PRL.62:349}.
The diffusion coefficient tends to minimal in the limits
$T\rightarrow 0$ and $T\rightarrow \infty$, indicating the crucial
role of thermal noise in the Brownian dynamics.
The maximum in the diffusion coefficient can be understood
by realizing that high temperature J\"{u}ttner distribution endows the
relativistic particle an effective mass $m_\textrm{eff}=\frac{k_BT}{v^2_\textrm{F}}$,
leading to a drop in diffusion at high temperature~\cite{Haenggi-RMP.81:387,Pototsky-EPJB.85:356}.
Moreover, comparison
of the blue solid, red asteric solid, green dash-dotted, and black dashed curves shows
that the diffusion coefficient is minimal for the twist
angle that matches the magic angle ($\theta = 1.05^{\circ}$) where the band structure is flat~\cite{Bistritzer-PNAS.108:12233,Koshino-PRX.8:031087}, corresponding
to minimum velocity of the particle and
consequently the diffusion rate is minimal. Hence, large diffusion is observed for $0<\theta < 1.05^{\circ}$.
Hence this mechanism appears in the presence of combined
action of thermal noise, spatially varying periodic ratchet potential,
and time-periodic modulating force fields.
\begin{figure}[!ht]
\centerline{\psfig{figure=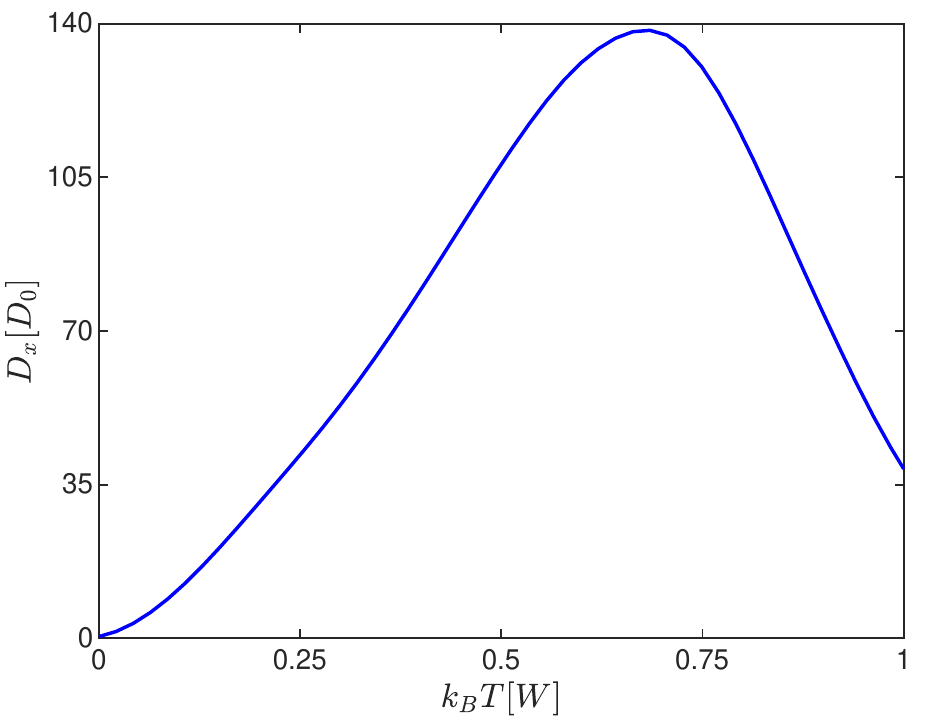,width=\columnwidth}}
\caption{Diffusion coefficient $D_x$
versus thermal noise strength $k_BT$ using $\theta = 1.05^{\circ}$
and $eE_x =0$ using the same parameters as used in Fig.~\ref{Figure3}.}
\label{Figure4}
\end{figure}
For simplicity and better understanding,
we analyze the diffusion in one dimension, i.e., $x$-direction and ac
drive forcing the charge carriers along $y$ with ${\bf{E}}\left(t\right)=\left(0,E_y\cos\left(\omega_yt\right)\right)$.
By numerically solving the semiclassical relativistic Langevin equations,
see Eqs.~\eqref{eq:LE1}-~\eqref{eq:LE4}, one can observe that a
relativistic particle tends to drift in the $x$-direction with
net current and diffusion coefficient.
More insight is obtained by plotting diffusion coefficient $D_x$ as
a function of thermal noise strength $k_BT$
using $\theta = 1.05^{\circ}$ as shown in Fig.~\ref{Figure4}.
This figure shows that diffusion coefficient is maximal around $k_BT\approx 0.65W$.
It is shown that a rectification is induced by $E_y(t)$ in
the $x$-direction at finite temperature.
The particle fluctuations are enhanced in the $x$-direction
under the influence of thermal noise around the minima
of periodic ratchet potential $V(x)$; as $k_y$ is driven by $E_y(t)$ toward values
on the order of $\sqrt{T}$, leading to maximize the
velocity $v_x$ and the Brownian
particle is kicked in the $x$-direction, either to the right
or to the left; this is how spatial asymmetry comes into play.
Interestingly, such a rectification mechanism becomes ineffective
when the particle sits at a potential minimum with $k_x=0$, which
only occurs in the absence of thermal noise, i.e., $T=0$.\\
\begin{figure}[!ht]
\centerline{\psfig{figure=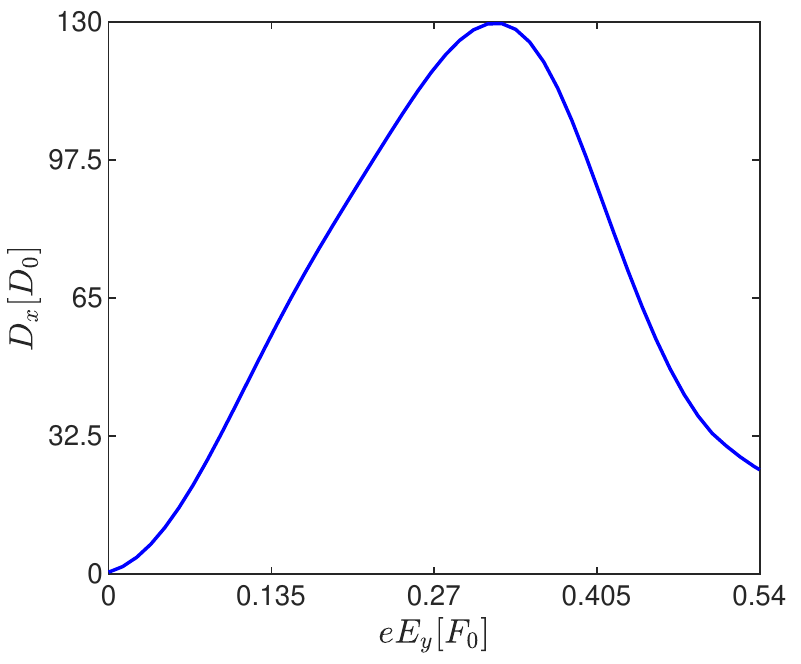,width=\columnwidth}}
\caption{Diffusion coefficient $D_x$ versus driving
field strength $eE_y$ using the same parameters as used for Fig.~\ref{Figure3}.}
\label{Figure5}
\end{figure}
Moreover, the diffusion of Brownian particle also depends
strongly on the amplitude of the driving force.
To demonstrate this effect, the diffusion coefficient $D_x$ is plotted as a function
of driving force amplitude $eE_y$ in Fig.~\ref{Figure5}. This figure
shows that the diffusion coefficient is maximal around $eE_y\approx 0.3F_0$,
where deterministically running solutions of the relativistic Langevin equations occur.
Interesting feature in Fig.~\ref{Figure5} is the
resonancelike behavior of the diffusion coefficient around an
optimal $eE_y$ for the enhancement of the diffusion rate.\\
It is known that there are two states of a driven Brownian dynamics:
locked state, in which the particle stays inside one potential well
that occurs in the regime of a small driving force strength, and
the running state, in which the particle runs over the
potential barriers which takes place when the amplitude of the external
field is large where both diffusive and regular behavior of the particle
dynamics can be observed. In these regions, the optimal matching of
the periodic force and thermal noise drives the charge carriers up the
potential hills during each time period. In turn these particles undergo
scattering at the potential barriers and finally diffuse rapidly
into the wide regions. It implies that an enhanced
diffusion rate is obtained under the optimal collective actions of the
spatial periodic potential, time periodic modulation, and stochastic stimulation.\\
\begin{figure}[!ht]
\centerline{\psfig{figure=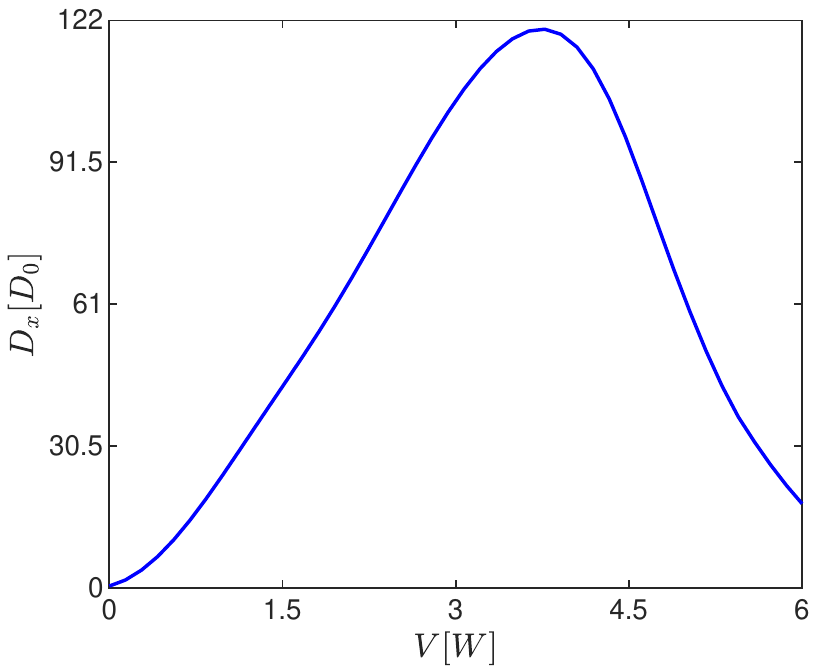,width=\columnwidth}}
\caption{Diffusion coefficient $D_x$ versus
periodic ratchet potential strength $V$ using the same
parameters as used for Fig.~\ref{Figure3}.}
\label{Figure6}
\end{figure}
Furthermore, the diffusion of Brownian particle is affected significantly
by changing strength of the ratchet potential of the substrate.
To realize this effect, the diffusion coefficient is plotted as a function
of ratchet potential strength in Fig.~\ref{Figure6}, showing
that the diffusion coefficient is maximal around $V\approx 3.7W$.
The interesting feature
of Fig.~\ref{Figure6} is the appearance of resonancelike
behavior of the diffusion coefficient around the optimal ratchet potential strength.
It mimics the behavior of the diffusion coefficient as
a function of thermal noise shown in Fig.~\ref{Figure4}.
If the potential wells are sufficiently deep relative to
thermal noise i.e. $V$ relative to $k_BT$ then diffusion is
suppressed as seen in Fig.~\ref{Figure6}.\\
\begin{figure}[!ht]
\centerline{\psfig{figure=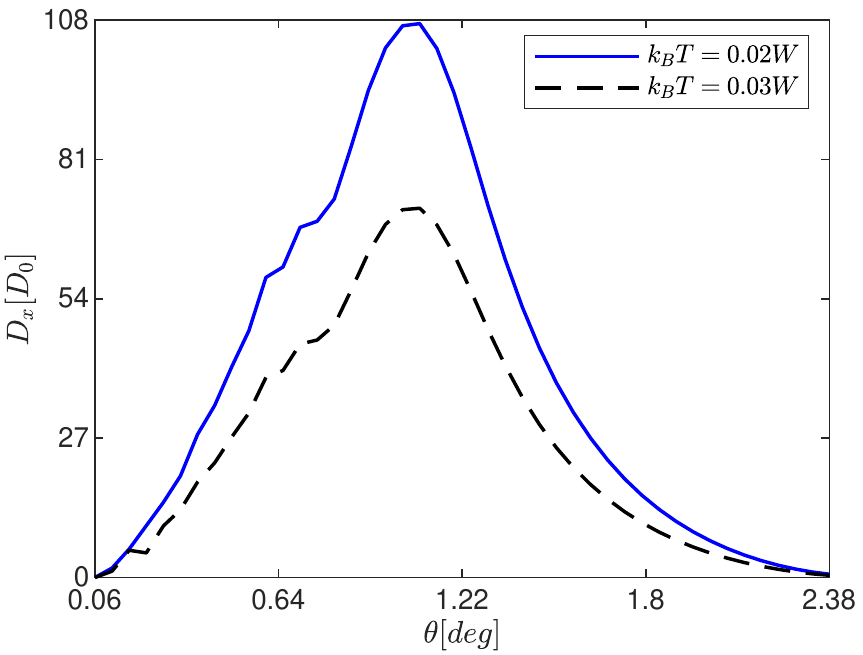,width=\columnwidth}}
\caption{Diffusion coefficient $D_x$ versus
periodic ratchet potential strength $V$ using the same
parameters as used for Fig.~\ref{Figure3}.}
\label{Figure7}
\end{figure}
Further understanding of the diffusion mechanism in twisted
bilayer graphene is obtained by analyzing the diffusion coefficient
versus twist angle as shown in Fig.~\ref{Figure7}.
It shows that the diffusion coefficient increases
with the increase of twist angle, hits maximum at $\theta \approx 1.03^{\circ}$
and then decreases rapidly to become minimal. Hence, it is maximal around the twist angle,
$\theta \approx 1.03^{\circ}$ which is in agreement with
the appearance of large diffusion coefficient for
$\theta = 1.03^{\circ}$ in Fig.~\ref{Figure3} and is minimal in the regions,
$\theta\ll 1.03^{\circ}$
and $\theta\gg 1.03^{\circ}$, illustrating the key
role of noise in the Brownian dynamics. This behavior
of the diffusion coefficient indicates the
existence of an optimal twist angle where the
diffusion is maximal. Moreover, comparison of
the blue solid and black dashed curves shows
that diffusion coefficient decreases with increase of thermal
energy of the system. Interestingly, Fig.~\ref{Figure7} also
shows that diffusion coefficient exhibits resonancelike
behavior around the value of twist angle that matches the optimal angle.
\subsubsection{Net current}\label{Sec:NC}
In this section, we consider charge carrier transport in
the presence of rectification
of non-equilibrium perturbations on a substrate
with periodic ratchet potential $V(x)$. It is
shown that for a nonrelativistic particle under the influence of two force
fields, the $x$ and $y$ components of position vector become
statistically uncorrelated, resulting into vanishing average
currents in both directions. However, for a relativistic particle the coupling
between $k_x$ and $k_y$ leads to mixing the orthogonal
components of ac electric field ${\bf{E}}$, producing a net current.
Hence drifting of a relativistic particle generates net current given by
\begin{align}\label{Eq:Current}
J=\frac{\braket{\dot{\bf{r}}}}{L}.
\end{align}
Eq.~\eqref{Eq:Current} reveals dependence of the net current on
both the driving frequency and temperature of the system.
\begin{figure}[!ht]
\centerline{\psfig{figure=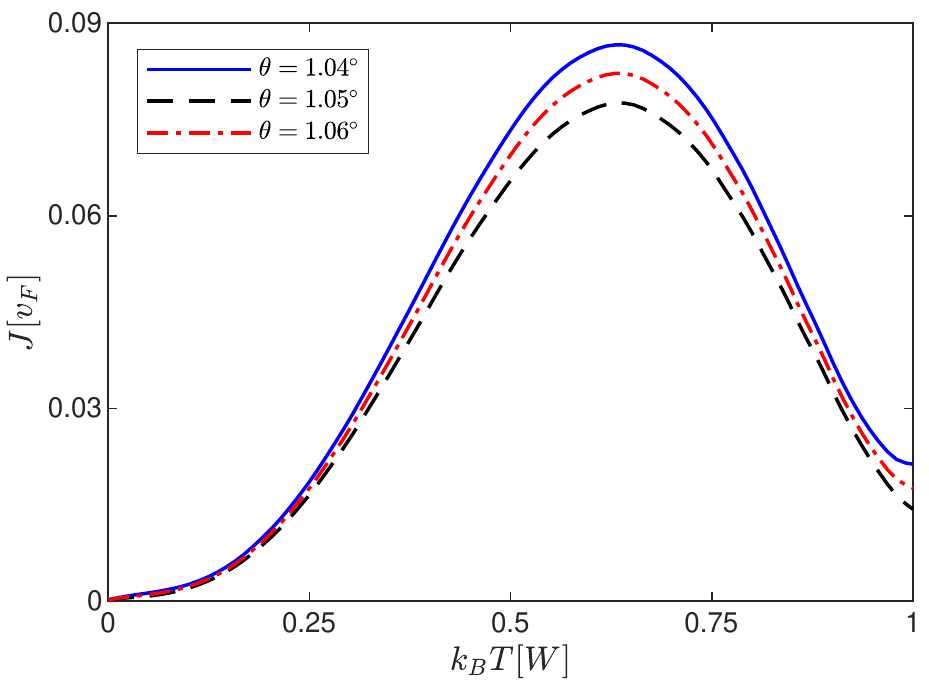,width=\columnwidth}}
\caption{Diffusion coefficient $D_x$ versus
periodic ratchet potential strength $V$ using the same
parameters as used for Fig.~\ref{Figure3}.}
\label{Figure8}
\end{figure}
The temperature dependence of this current is plotted in
Fig.~\ref{Figure8} for different values of the twist angle. It is evident from inspection
of this figure that the current initially increases with increasing
temperature, becomes maximal around $0.65W$ and then decreases. This behavior
of the net current illustrates the significant role played
by noise in the Brownian dynamics.
Comparison of the
blue solid, red dash-dotted, and black dashed curves shows that the net
current in twisted bilayer graphene depends strongly on the twist angle,
where the net current is minimal for the magic angle, $\theta = 1.05^{\circ}$
which is in agreement with the results for diffusion coefficient in Fig.~\ref{Figure3}.\\
It is shown that the results of the present study
are significantly different from the previous studies~\cite{Pototsky-EPJB.85:356,Reiman-PRE.65:031104,Bandyopadhyay-PRE.73:051108} due to the
use of totally different model, encoding completely different
effects and the underlying mechanism in this study. In particular,
our results are different from the results of Ref.~\cite{Pototsky-EPJB.85:356}.
In the present work, the relativistic
ratcheting depends strongly on the twist angle
which is not relevant to monolayer graphene.
In addition, we have analyzed the trajectories of Brownian
particle and diffusion coefficient as a function of ratchet
potential strength, driving field amplitude, and twist angle which are
missing in the aforementioned paper. Moreover,
the resonancelike behaviors of the diffusion coefficient and net
current in twisted bilayer graphene are well pronounced compared to monolayer graphene.
This difference appears due to the peculiar band structure of
TBG that can be tuned through the twist angle that in turn
affects significantly the dynamics of charge carriers
in the system, presenting different scenario from monolayer graphene.\\
Finally, we propose experimental measurement of the
Brownian dynamics of Dirac fermions in
TBG using the recently developed experimental techniques~\cite{Kheifets-Sc.343:1493,Petrov-SR.12:8618,Tamtogl-NC.11:278,Grimm-PRE.86:021912}.
\section{Conclusions}\label{Sec:Conc} 
In summary, Brownian dynamics of Dirac fermions in twisted
bilayer graphene under the influence of orthogonal, commensurate
ac drives in a periodic ratchet potential of a substrate has been
investigated, illustrating key role of the twist angle.
We have found that the real space trajectories are changed
significantly by changing the twist angle.
The diffusion coefficient exhibits resonancelike behavior around the
optimal thermal noise strength, i.e., $k_BT \sim 0.65W$, where diffusion rate is maximal.
The diffusion coefficient tends to minimal in the limits
$T\rightarrow 0$ and $T\rightarrow \infty$, indicating the crucial
role of thermal noise in the Brownian dynamics. Similar behaviors
are exhibited by the diffusion coefficient as a function of ratchet
potential strength and driving force field amplitude.
It has been shown that the diffusion coefficient is minimal for twist
angle that matches the magic angle ($\theta = 1.05^{\circ}$),
whereas large diffusion coefficient is observed for $0<\theta < 1.05^{\circ}$.
Moreover, we have shown that the Dirac fermions in TBG exhibit
remarkable ratchet effect as a net current.
Analysis reveals that the net current initially increases with increase
in temperature, becomes maximal and then decreases rapidly, illustrating
the significant role played by thermal noise in the dynamics.
As a consequence, as threshold parameter matches
its optimal value, deterministic
running states appear in the driven Brownian dynamics in the
limit of weak thermal noise where the diffusion enhances
significantly compared to bare thermal diffusion.
We expect experimental measurement of the Brownian motion of Dirac
fermions in TBG using the recently developed experimental techniques~\cite{Kheifets-Sc.343:1493,Petrov-SR.12:8618,Tamtogl-NC.11:278,Grimm-PRE.86:021912}.\\
Finally, it is illustrated that the findings of this work can be
utilized in realizing electronic devices based on graphene cutting edge technologies.
\section*{Acknowledgments}
The author cordially acknowledges the support of Higher Education Commission (HEC),
Pakistan under National Research Program for Universities NRPU Project No. 11459.

\section*{Data availability statement}
Data sharing is not applicable to this article, as it describes
entirely theoretical research work.
\section*{Declaration of competing interest}
The author declares that he has no known competing financial
interests or personal relationships that could have appeared to
influence the work reported in this paper.

\end{document}